\documentclass[a4paper]{article}  
\usepackage{indentfirst}  
\usepackage{bm}    
\usepackage{graphicx}  
\usepackage{graphicx}
\usepackage{enumitem}
\usepackage{subfigure}
\usepackage{multirow}
\usepackage[table,xcdraw]{xcolor}
\usepackage{url}
\usepackage{fancyhdr}

\usepackage{geometry}
\begin{document}    
\thispagestyle{empty} \vspace*{0.8cm}\hbox
to\textwidth{\vbox{\hfill\noindent \\ \textit{Proceedings of the 8th International Conference on Pedestrian and Evacuation Dynamics (PED2016)\\
Hefei, China - Oct 17 -- 21, 2016\\
Paper No. 41}
\hfill}}
\par\noindent\rule[3mm]{\textwidth}{0.2pt}\hspace*{-\textwidth}\noindent
\rule[2.5mm]{\textwidth}{0.2pt}

\begin{center}
\LARGE\bf Social Groups and Pedestrian Crowds: Experiment on Dyads in a Counter Flow Scenario
\end{center}

\begin{center}
\rm Andrea Gorrini$^{1}$\footnote{Corresponding author.} \ Luca Crociani$^1$ \ Claudio Feliciani$^2$ \ Pengfei Zhao$^3$ \\ Katsuhiro Nishinari$^4$ and Stefania Bandini$^{1,4}$
\end{center}

\begin{center}
\begin{small} \sl
${}^{\rm 1}$ Complex Systems and Artificial Intelligence research center, University of Milano-Bicocca \\Viale Sarca 336 - Ed. U14, 20126 Milano (ITALY) \\   
andrea.gorrini@unimib.it; luca.crociani@disco.unimib.it; stefania.bandini@disco.unimib.it \\
${}^{\rm 2}$ Department of Advanced Interdisciplinary Studies, Graduate School of Engineering, The University of Tokyo\\ 4-6-1 Komaba, Meguro-ku, Tokyo 153-8904 (JAPAN)\\
feliciani@jamology.rcast.u-tokyo.ac.jp\\
${}^{\rm 3}$ Beijing University of Technology \\ Jiaotong Building 205, Pingleyuan 100, Chaoyang District, Beijing (CHINA)\\
zhao\_pf@emails.bjut.edu.cn\\
${}^{\rm 4}$ Research Center for Advanced Science and Technology, The University of Tokyo \\4-6-1 Komaba, Meguro-ku, Tokyo 153-8904 (JAPAN) \\ 
tknishi@mail.ecc.u-tokyo.ac.jp\\
\end{small}
\end{center}
\vspace*{2mm}

\begin{center}
\begin{minipage}{15.5cm}
\parindent 20pt\small
\noindent\textbf{Abstract -} The calibration and validation of pedestrian simulations require the acquisition of empirical evidences of human behaviour. The current work presents the results of an experiment focused on the potentially combined effect of counter flow and grouping on pedestrian dynamics. In particular, we focused on: (\emph{i}) four different configurations of flow ratio (the rate between the minor flow and the total flow in bidirectional scenarios);  (\emph{ii}) dyads, as the most frequently observed and basic social groups of crowds. Results showed that the increase of flow ratio negatively impacted the speed of pedestrians. Dyads walked significantly slower than singletons, due to the difficulty in movement coordination among group members (proxemics) in case of counter flow. The collected results represent an useful contribution towards the validation of pedestrian simulations.
\end{minipage}
\end{center}

\begin{center}
\begin{minipage}{15.5cm}
\begin{minipage}[t]{2.3cm}{\bf Keywords:}\end{minipage}
\begin{minipage}[t]{13.1cm}
Pedestrian, Experiment, Counter Flow, Bidirectional Flow, Dyads, Proxemics 
\end{minipage}\par\vglue8pt
\end{minipage}
\end{center}

\section{Introduction}  

The role of advanced computer-based systems for the simulation of pedestrian dynamics is becoming a consolidated and successful field of research and application, thanks to the possibility to test the efficiency, comfort and safety of urban crowded facilities (e.g., travel time, queuing, level of service, walkability). The development of simulation systems requires a cross-disciplinary methodology in order to calibrate and validate the model according to empirical evidences  about pedestrian behaviours. In this framework, the objective of the current experimental study is to analyse the potentially combined impact of counter flow ratio and grouping (\emph{dyads}) on pedestrian crowd dynamics, for sake of model validation.

Counter flow phenomenon has been largely studied in the literature, however, the effects of bidirectional flows on pedestrian dynamics are not yet completely understood. Some authors \cite{kretz2006experimental,lam2002study} found that balanced bidirectional streams perform better than unidirectional ones, while different authors showed that this phenomenon negatively affects the walking movements, with turbulences and competitive interactions among pedestrians \cite{zhang2012ordering,alhajyaseen2011effects,feliciani2015phenomenological}.

Grouping phenomenon in pedestrian crowds has been generally neglected until the last decade in the literature of simulations, in which microscopic modelling of individual behaviour has constituted the main focus of this research field \cite{templeton2015mindless}. However, recent contributions \cite{moussaid2010walking,zanlungo2014potential} brought much interests on this element and approaches considering its presence in the simulations have been proposed \cite{vizzari2013adaptive,muller2014study}.

Urban pedestrian facilities are characterised indeed by the presence of groups \cite{willis2004human,Schultz2008}: social units featured by common goals and variable strength of membership, generally moving slower than single pedestrians. In particular, the granulometric distribution of pedestrian flows is strongly affected by two-members groups, which represent the most frequently observed and basic social groups of crowds \cite{CD_AndreaGorrini2016}. Analyses of pedestrian dynamics not considering this aspect have a reduced accuracy, since grouping was found to change crowd movements and evacuation dynamics \cite{muller2014study}. 

In particular, elements such as the level of density, the presence of obstacles and the geometry of the environment, can create difficulties in movement coordination among group members, depending on the need to maintain spatial cohesion to communicate while walking (\emph{proxemics}) \cite{CostaGroupDistances}, generating different spatial patterns at variable density conditions: line abreast pattern at low density, diagonal pattern at medium density, river-like pattern at high density. The experiment proposed in this paper (Sec. 2) is aimed at testing the combined effect of counter flow ratio and dyads on pedestrian dynamics, focusing on speeds, trajectories and proxemics (Sec. 3). 

\section{Experimental Study}
\label{sec:exp}
\begin{figure}[t!]
\begin{center}
\subfigure[Procedure No. 1 - Flow Ratio = 0]{\includegraphics[width=.4\textwidth]{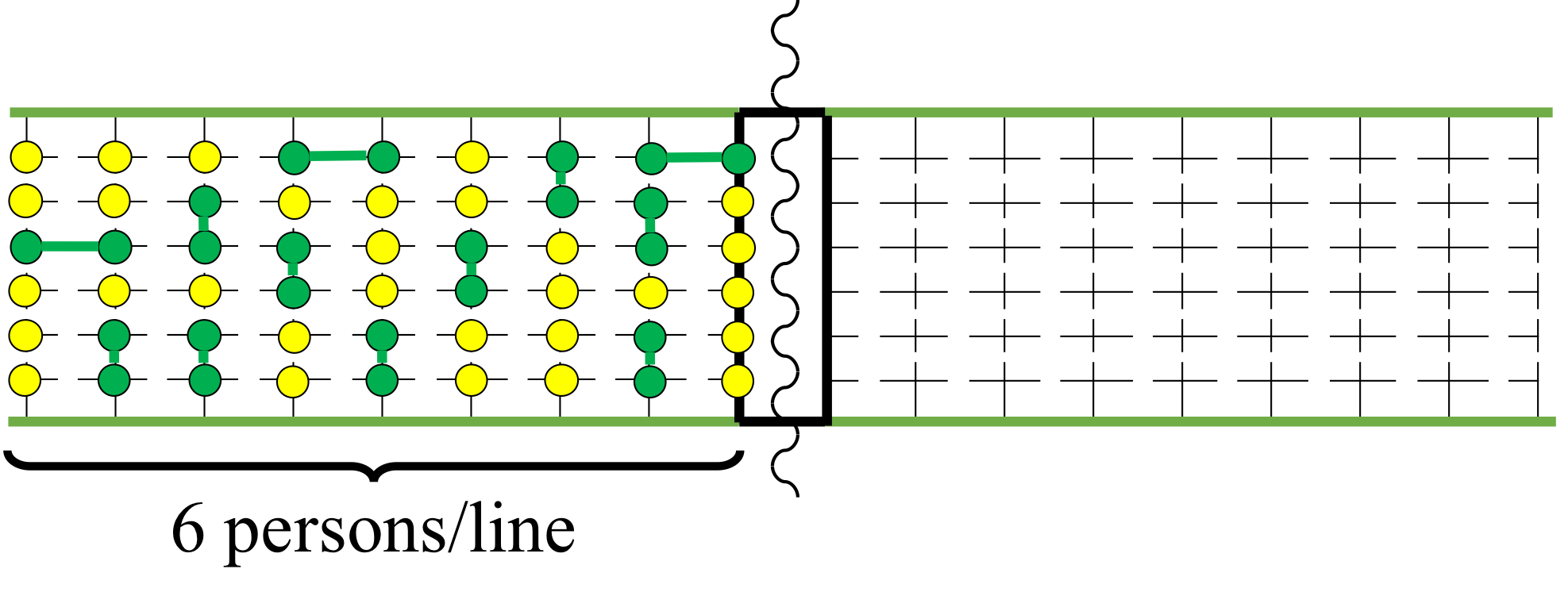}} \hspace{1cm}
\subfigure[Procedure No. 2 - Flow Ratio = .167]{\includegraphics[width=.4\textwidth]{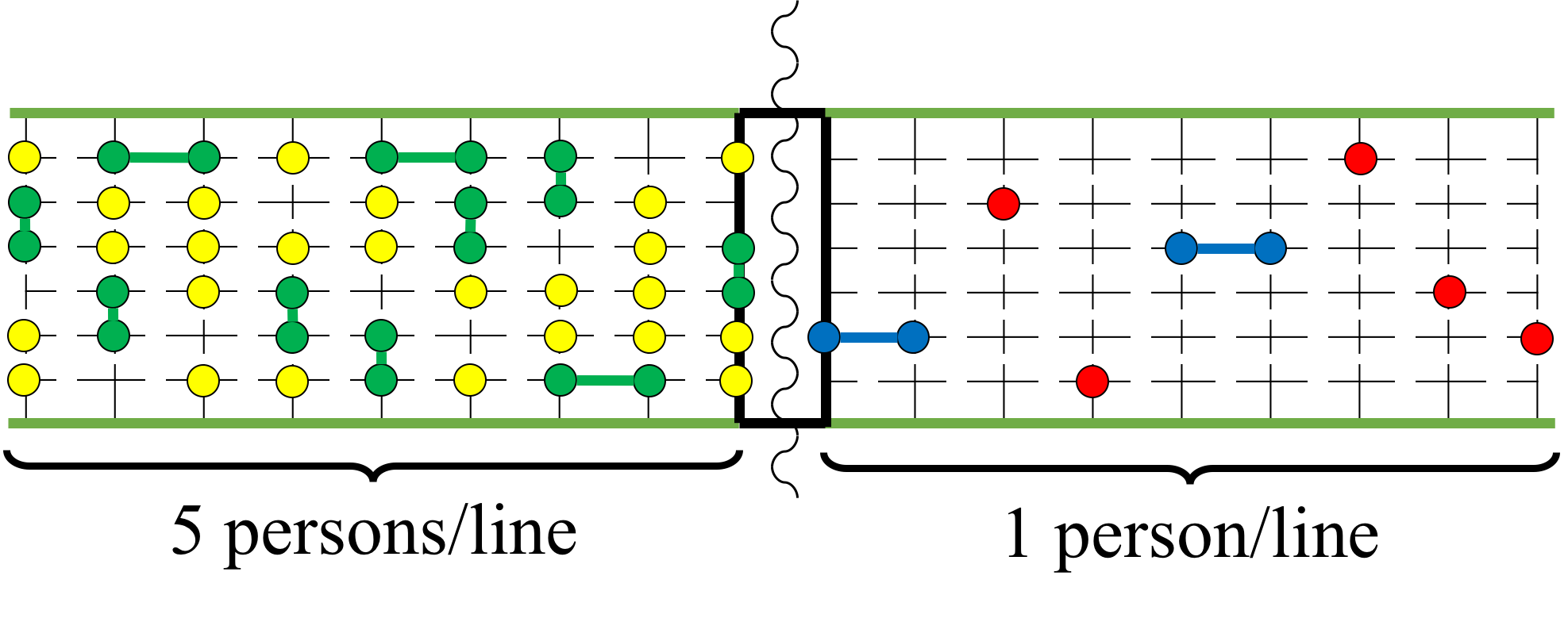}}
\subfigure[Procedure No. 3 - Flow Ratio = .333]{\includegraphics[width=.4\textwidth]{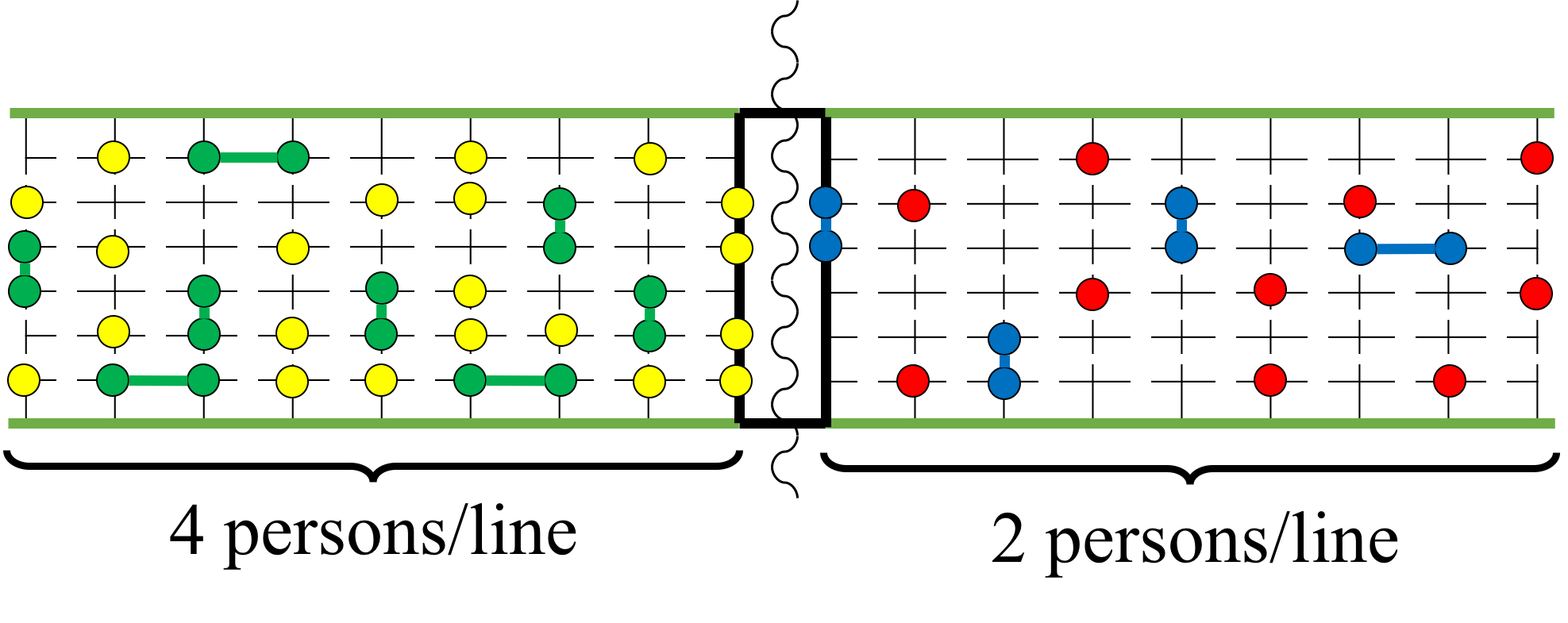}} \hspace{1cm}
\subfigure[Procedure No. 4 - Flow Ratio = .5]{\includegraphics[width=.4\textwidth]{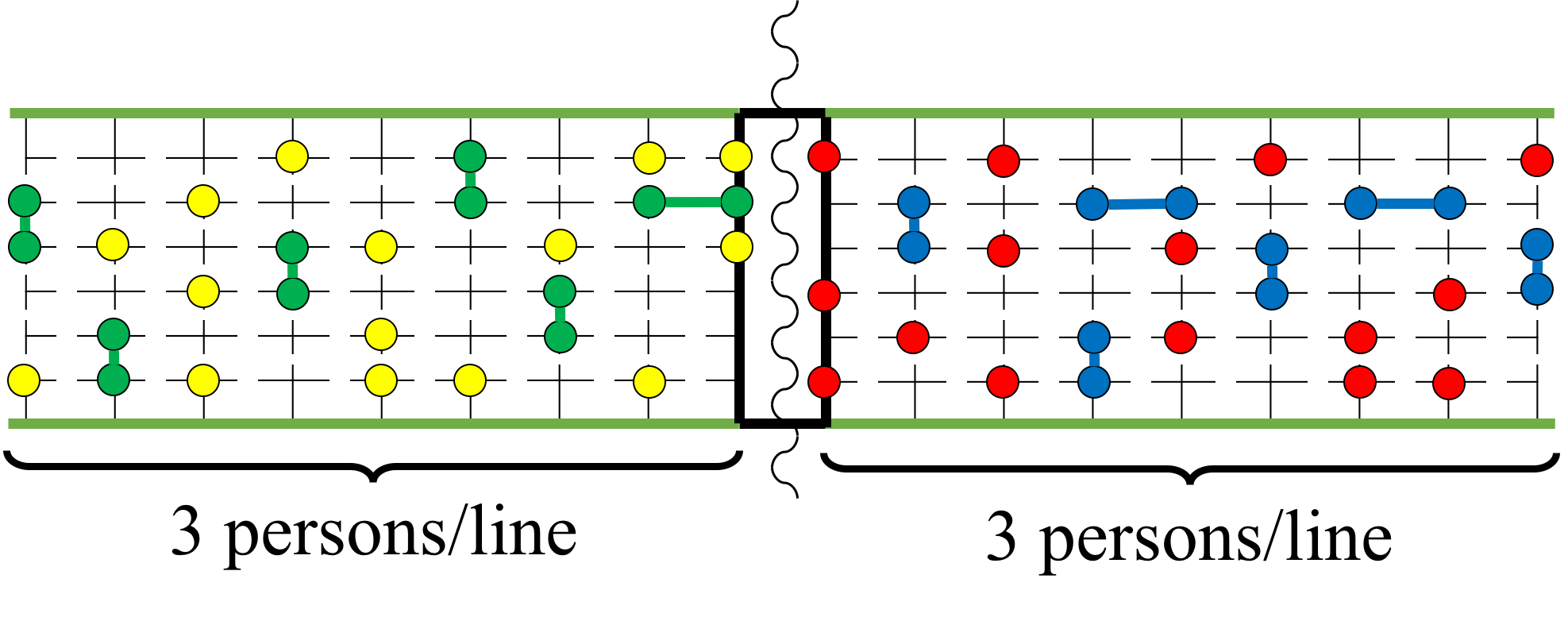}}
\caption{The starting positions of participants. Linked dots represent dyads. Colours refer to the hats used for tracking, identifying the type of participants (flow direction, singles and dyads).}
\label{fig:scheme}
\end{center}
\end{figure}

The experiment was performed on June 13, 2015 at the Research Center for Advanced Science and Technology of The University of Tokyo (Tokyo, JAPAN). The experiment was aimed at testing the following hypotheses: (Hp1) the increase of flow ratio negatively impacts the speed of pedestrians; (Hp2) dyads walk slower than singles, due to need to maintain spatial cohesion. In this paper we denote the \emph{flow ratio} as the rate between the minor flow and the total flow in bidirectional scenarios. Flow ratio was managed as independent variable among four different experimental procedures (see Fig. \ref{fig:scheme}):

\begin{enumerate}
\item Flow ratio = 0 (unidirectional flow) $\rightarrow$ 6 persons per line, 30 singles, 24 dyad members;
\item Flow ratio = .167 $\rightarrow$ minor flow 27\%, 1 person/line, 5 singles, 4 dyad members; major flow 73\%, 5 persons/line, 25 singles, 20 dyad members;
\item Flow ratio = .333 $\rightarrow$ minor flow 33\%, 2 persons/line, 10 singles, 8 dyad members; major flow 67\%, 4 persons/line, 21 singles, 16 dyad members;
\item Flow ratio = .5 (balanced counter flow) $\rightarrow$ right to left flow 50\%, 3 persons/line, 15 singles, 12 dyad members; left to right flow 50\%, 3 persons/line, 15 singles, 12 dyad members.
\end{enumerate}

The setting was designed as a corridor-like scenario delimited with band partitions. It was composed of: (\emph{i}) a measurement area in the centre (10 m $\times$ 3 m); (\emph{ii}) two side buffer zones to allow pedestrians reaching a stable speed (2 m $\times$ 3 m); (\emph{iii}) two starting areas (12 m $\times$ 3 m). The starting positions were drawn on the floor with coloured signs (see Fig. \ref{fig:scheme}): lateral distance 0.5 m, longitudinal distance 1.5 m. The number of dyads per line and the relative positions of group members were homogeneously distributed. At the start signal, all participants were asked to walk toward the opposite side of the corridor. Each procedure was repeated four times, asking participants to change their starting positions.

The sample was composed of 54 participants (from 18 to 25 years old), Japanese male students of The University of Tokyo. 24 participants were randomly paired (44\% of the total, as observed in \cite{CD_AndreaGorrini2016}) and asked to walk close to each other during the experiment reproducing the proxemic  behaviour of social groups (no instruction was given to them about the spatial pattern to maintain while walking). The rest of participants walked as individual pedestrians. 

\begin{figure}[t!]
\begin{center}
\subfigure[Procedure No. 1 - Flow Ratio = 0]{\includegraphics[width=.4\textwidth]{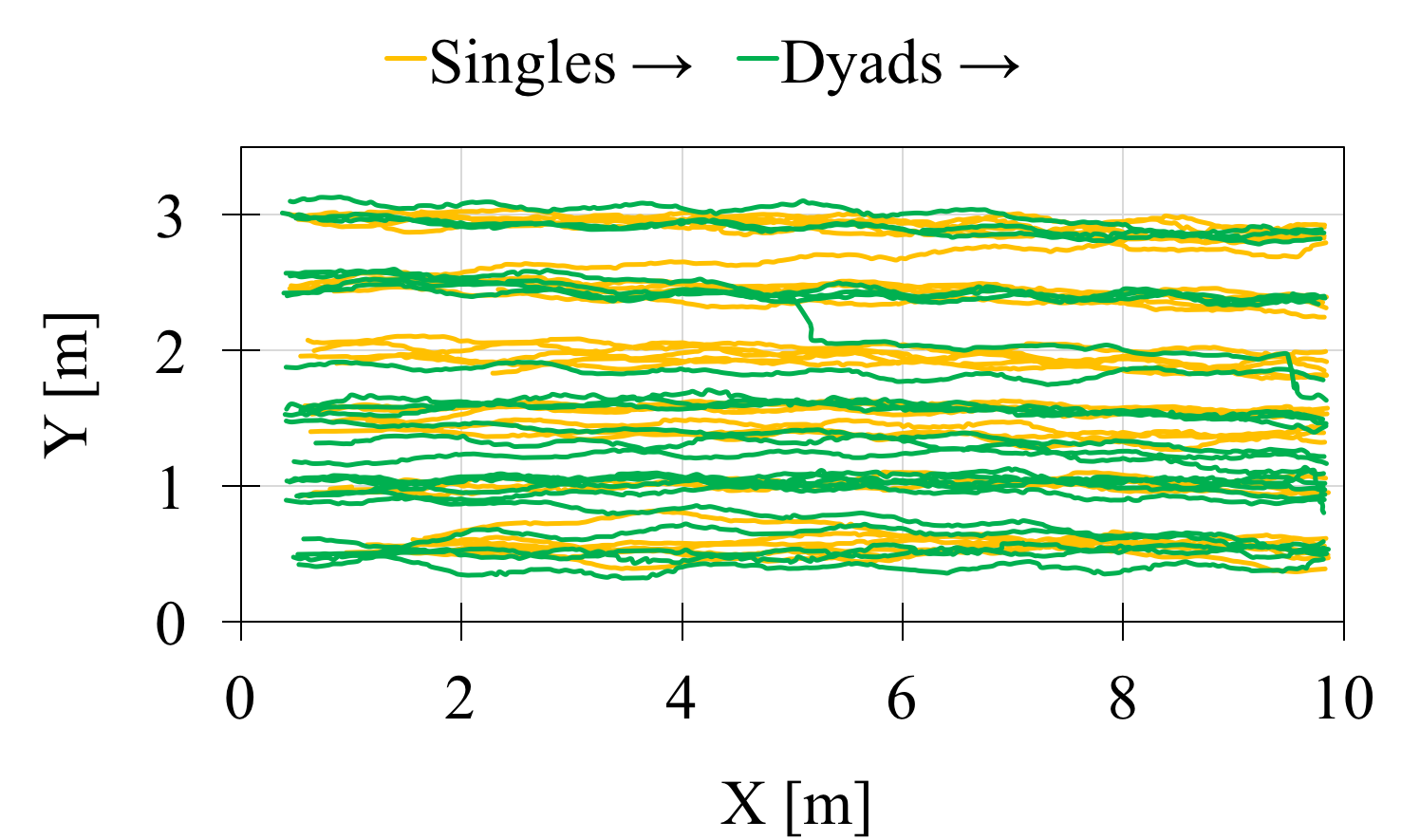}}\hspace{1cm}
\subfigure[Procedure No. 2 - Flow Ratio = .167]{\includegraphics[width=.4\textwidth]{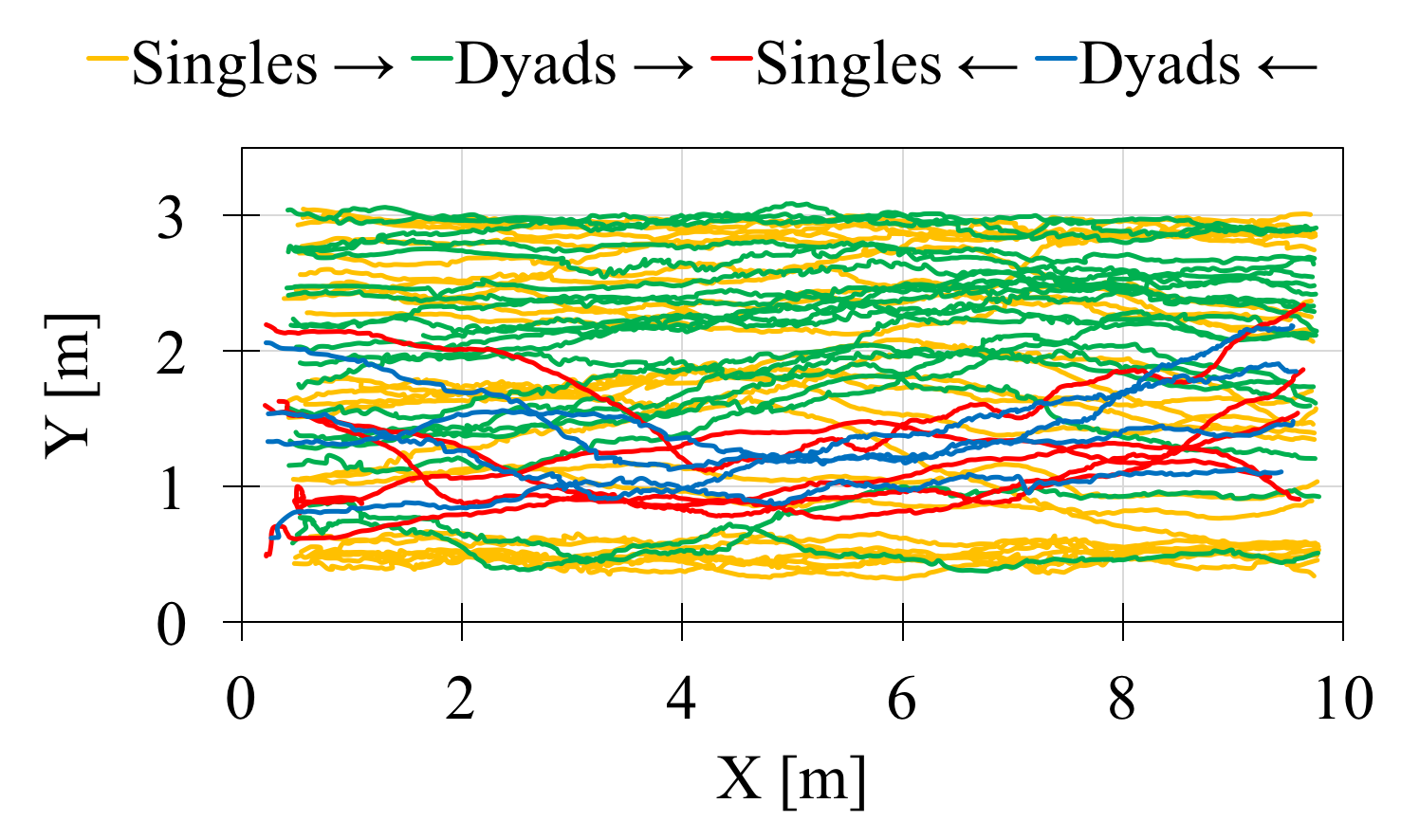}}
\subfigure[Procedure No. 3 - Flow Ratio = .333]{\includegraphics[width=.4\textwidth]{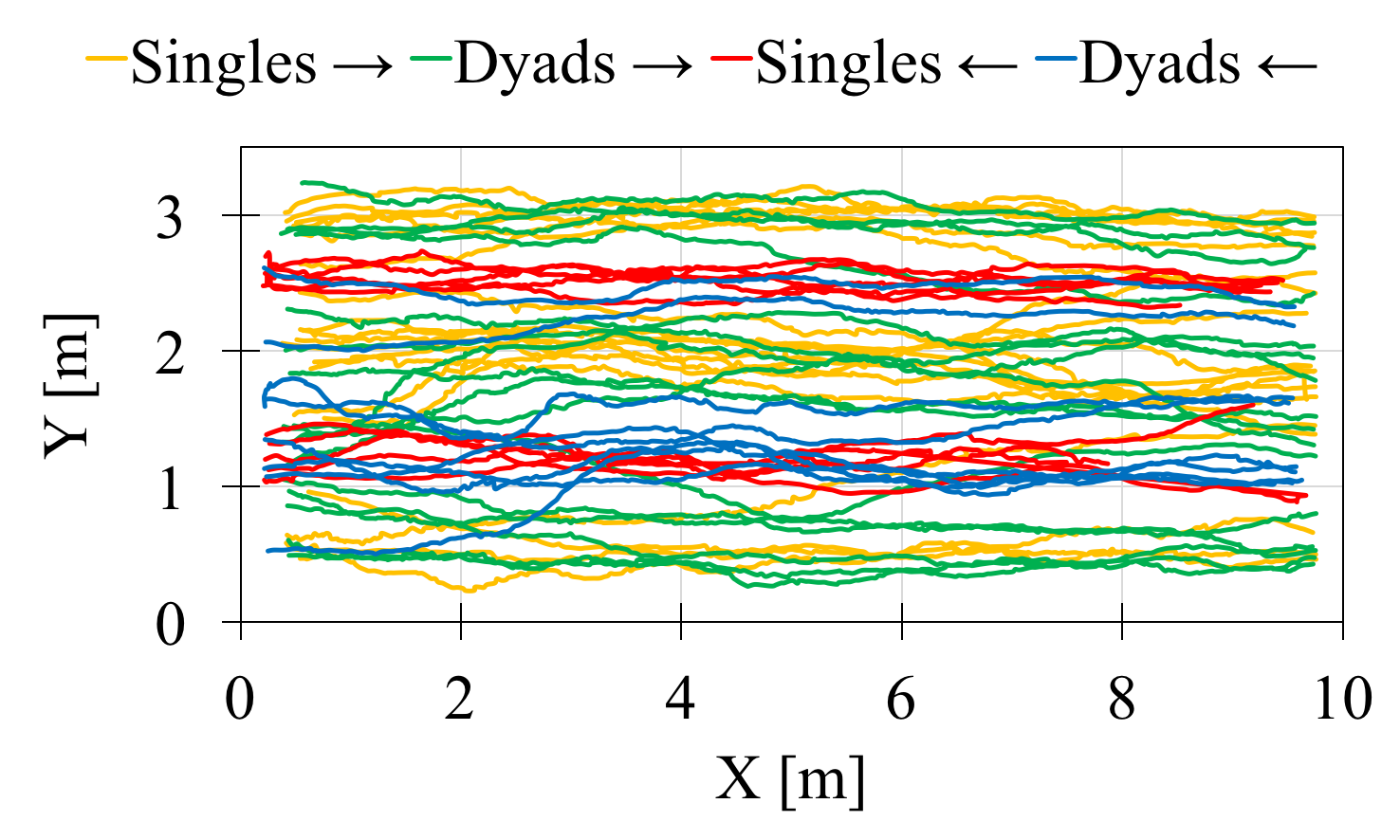}}\hspace{1cm}
\subfigure[Procedure No. 4 - Flow Ratio = .5]{\includegraphics[width=.4\textwidth]{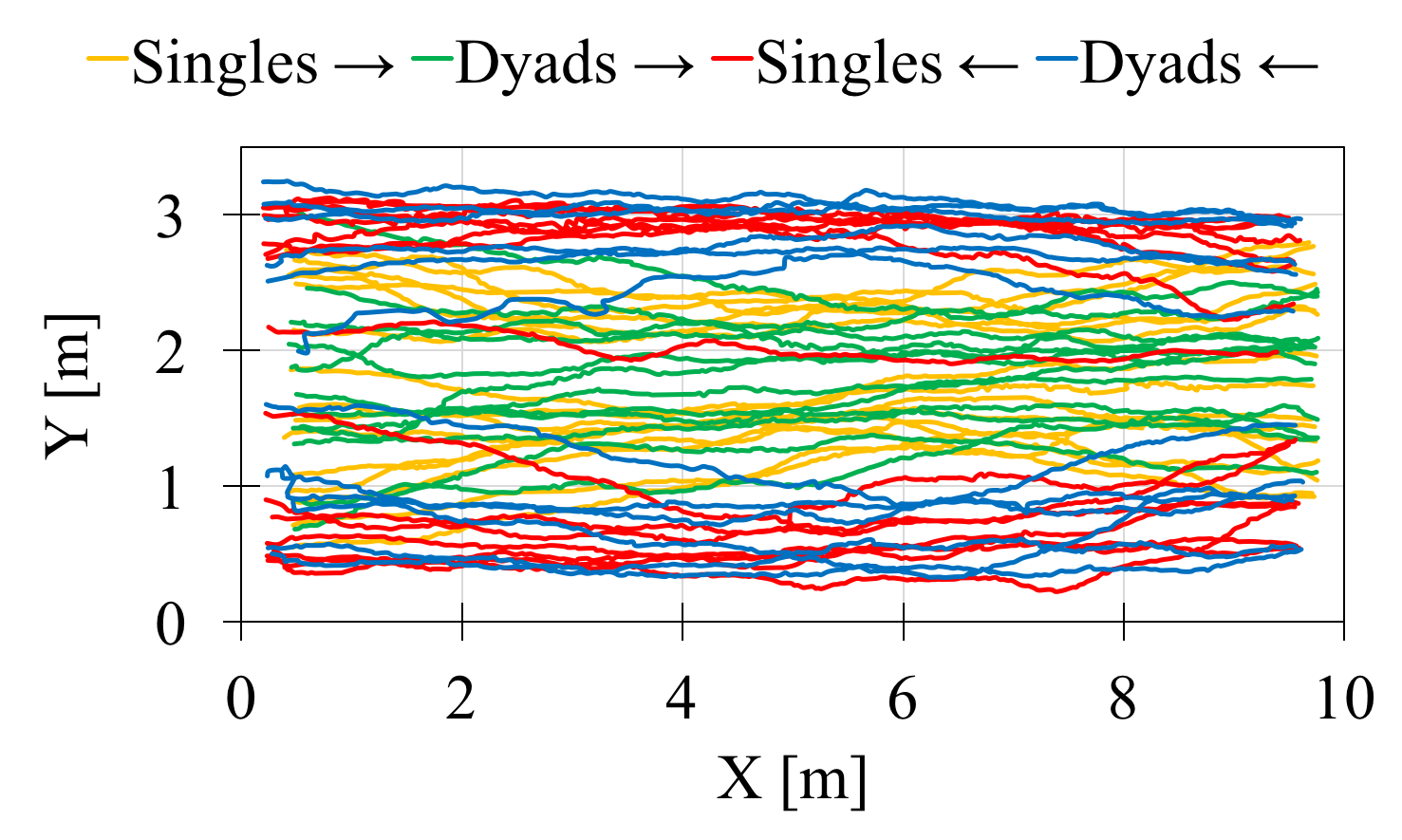}}
\caption{The trajectories of singles and dyad members.}
\label{fig:tr}
\end{center}
\end{figure}

\begin{figure}[t!]
\begin{center}
\includegraphics[width=.8\textwidth]{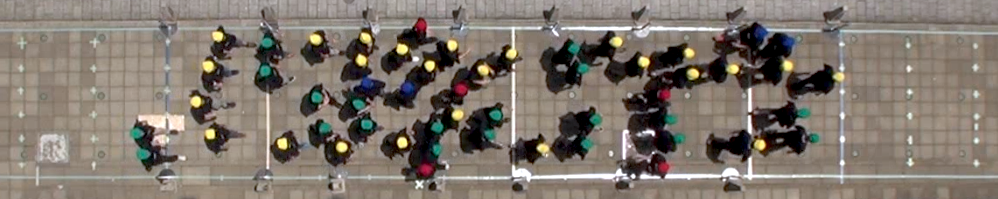}
\caption{A video frame of the analysed measurement area of the experiment (procedure No. 2).}
\label{fig:exp_frame}
\end{center}
\end{figure}

An HD wide-lens camera was located from a zenith position at an height of 21 m in order to record the movements of the pedestrians within the measurement area. All participants wore black T-shirts and caps with different colours depending on the flow direction and singles/dyads subsamples. The recorded video images have been firstly crop and corrected to avoid the image distortion due to the wide lens, and then analysed by using the PeTrack software \cite{boltes2013collecting}, which allowed to automatically track pedestrians\rq\ trajectories. A screenshot from the video footages of the experiment is shown in Fig. \ref{fig:exp_frame}. The trajectories of participants, one iteration per procedure, are shown in Fig. \ref{fig:tr}. 

\section{Analysis of Flow Ratio, Speed and Proxemics}
\label{sec:exp_results}

\begin{figure}[t!]
\begin{center}
\subfigure{\includegraphics[width=.5\textwidth]{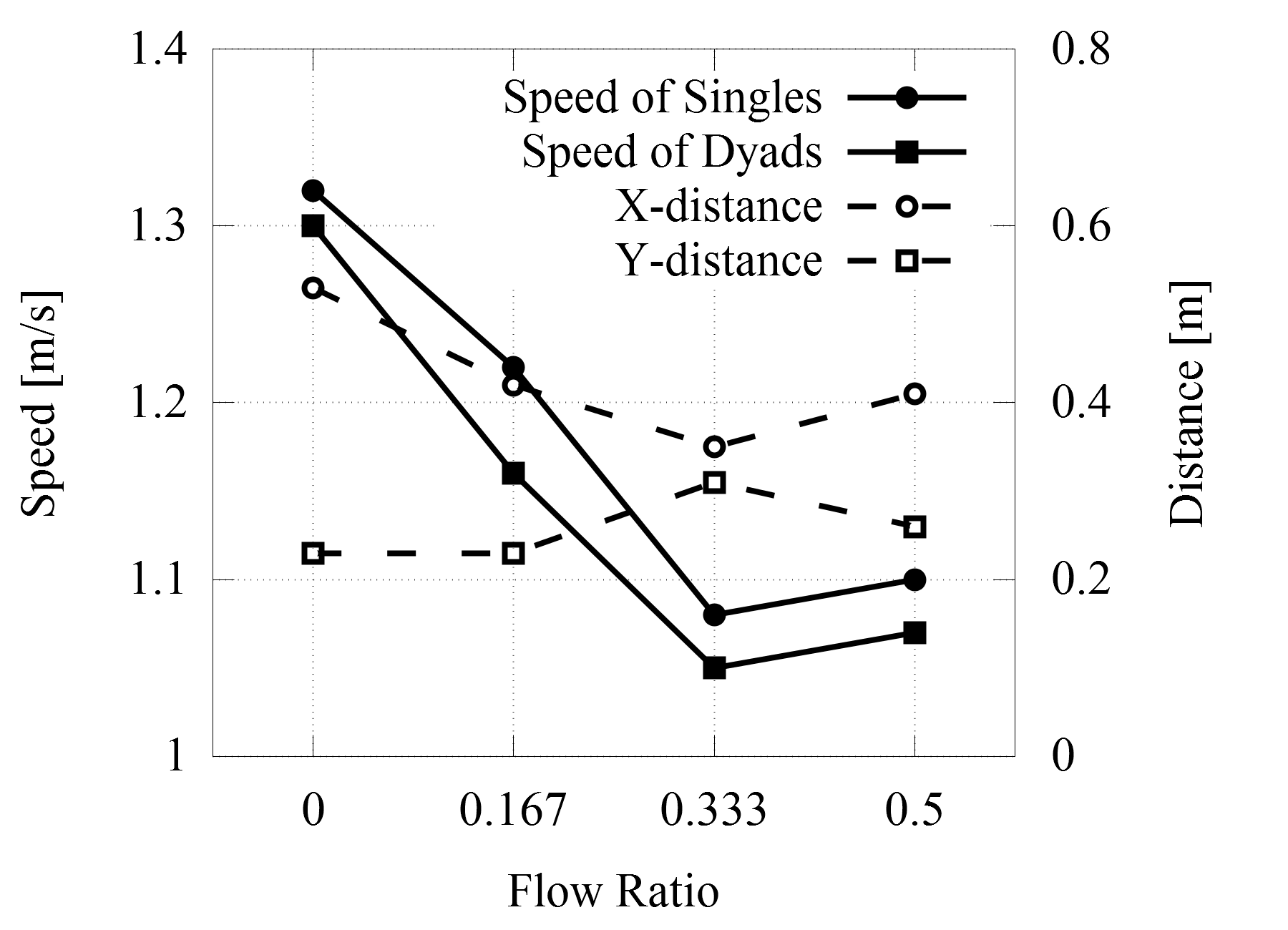}}
\caption{Results about speed of singles and dyads, X and Y-distance between group members.}
\label{fig:comparison}
\end{center}
\end{figure}

\begin{table}[t!]
\centering
\caption{Mean and standard deviation of results.}\vspace{.5cm}
\label{tab:speed}
\small
\begin{tabular}{|l|l|l|l|l|l|}
\hline
\multicolumn{1}{|l|}{\cellcolor[HTML]{EFEFEF}EXP Procedures} & \multicolumn{1}{l|}{\cellcolor[HTML]{EFEFEF}Flow Ratio} & \cellcolor[HTML]{EFEFEF}Grouping & \cellcolor[HTML]{EFEFEF}Speed {[}m/s{]} & \cellcolor[HTML]{EFEFEF}X-distance {[}m{]} & \cellcolor[HTML]{EFEFEF}Y-distance {[}m{]} \\ \hline\hline
 &  & Total & 1.31 $\pm$ .10 SD & - & - \\ \cline{3-6} 
 &  & Singles & 1.32 $\pm$ .11 SD & - & - \\ \cline{3-6} 
\multirow{-3}{*}{No. 1} & \multirow{-3}{*}{0} & Dyads & 1.30 $\pm$ .09 SD & .53 $\pm$ .21 SD & .23 $\pm$ .25 SD \\ \hline 
 &  & Total & 1.20 $\pm$ .13 SD & - & - \\ \cline{3-6} 
 &  & Singles & 1.22 $\pm$ .15 SD & - & - \\ \cline{3-6} 
\multirow{-3}{*}{No. 2} & \multirow{-3}{*}{.167} & Dyads & 1.16 $\pm$ .10 SD & .42 $\pm$ .13 SD & .23 $\pm$ .24 SD \\ \hline 
 &  & Total & 1.07 $\pm$ .09 SD & - & - \\ \cline{3-6} 
 &  & Singles & 1.08 $\pm$ .10 SD & - & - \\ \cline{3-6} 
\multirow{-3}{*}{No. 3} & \multirow{-3}{*}{.333} & Dyads & 1.05 $\pm$ .09 SD & .35 $\pm$ .17 SD & .31 $\pm$ .22 SD \\ \hline 
 &  & Total & 1.08 $\pm$ .10 SD & - & - \\ \cline{3-6} 
 &  & Singles & 1.10 $\pm$ .10 SD & - & - \\ \cline{3-6} 
\multirow{-3}{*}{No. 4} & \multirow{-3}{*}{.5} & Dyads & 1.07 $\pm$ .11 SD & .41 $\pm$ .13 SD & .26 $\pm$ .20 SD \\ \hline
\end{tabular}
\end{table}

A two-factors analysis of variance\footnote{Statistics have been conducted at the p $<$ .01 level, relying on repeated measures (4 iterations per procedure).} (two-way ANOVA) was conducted to test the potentially combined effect of \emph{flow ratio} and \emph{grouping} on speed (see Tab. \ref{tab:speed} and Fig. \ref{fig:comparison}). Results showed a significant effect for the flow ratio factor [F(3,856) = 242.777, p $<$ .000], and a significant effect for the grouping factor [F(1,856) = 26.946, p $<$ .000]. No significant interaction among the two factors was found [F(3,856) = 1.008, p = .388]. 

A linear regression showed that the overall speed of participants was affected by the increase of flow ratio [F(1,862) = 546.039, p $<$ .000, R-square of .388; speed =  1.287 - .489 * flow ratio]. A post hoc Tukey test showed that the speed of participants among the procedures No. 3 (flow ratio = .333) and No. 4 (flow ratio = .5) did not differ significantly at p = .406. A post hoc Tukey test showed that the speed of singles and dyads among the procedures No. 1 (flow ratio = 0) did not differ significantly at p = .05056. The average speed of singles (1.13 m/s $\pm$ .08) and dyads (1.09 m/s $\pm$ .06) among procedures No. 2, No. 3 and No. 4 differed significantly at p $<$ .000. 

\begin{figure}[t!]
\begin{center}
\subfigure[ Procedure No. 1 - Flow Ratio = 0]{\includegraphics[width=.43\textwidth]{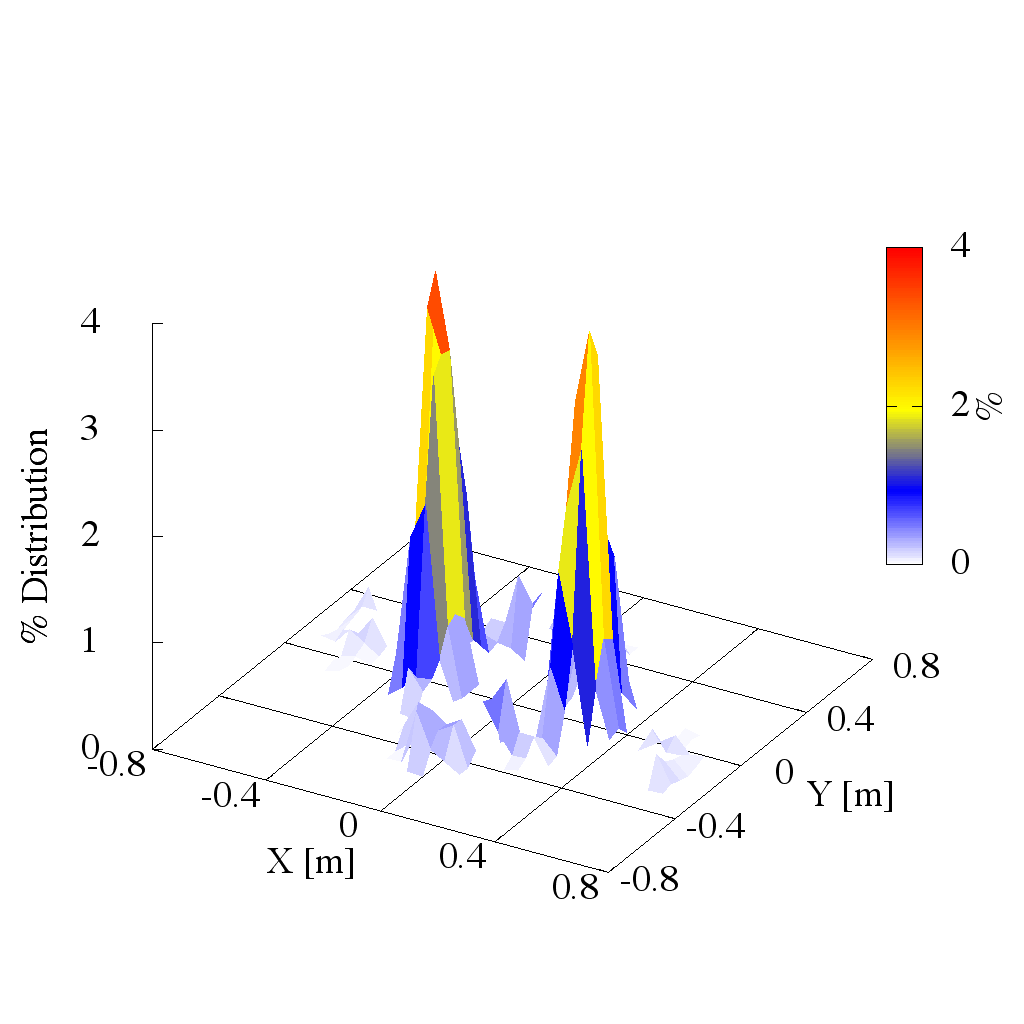}}
\subfigure[ Procedure No. 2 - Flow Ratio = .167]{\includegraphics[width=.43\textwidth]{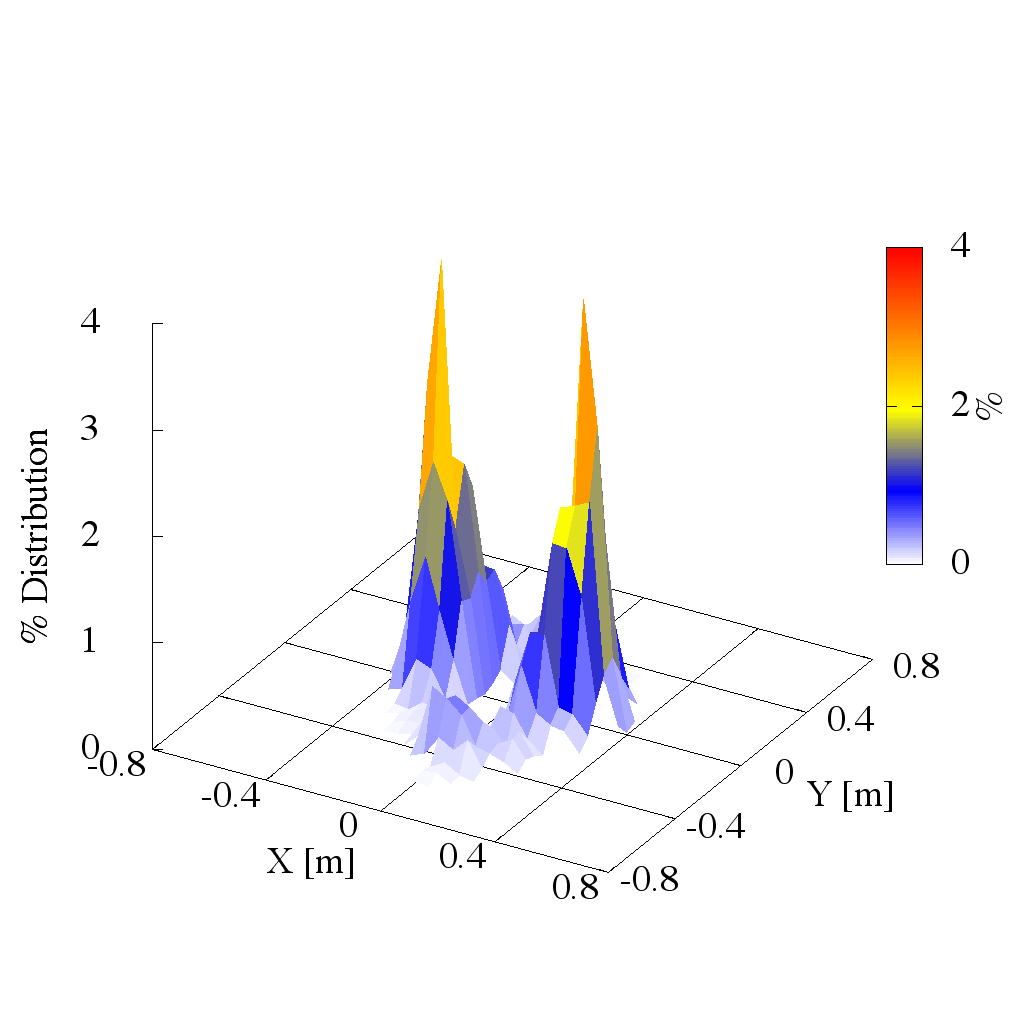}}
\subfigure[ Procedure No. 3 - Flow Ratio = .333]{\includegraphics[width=.43\textwidth]{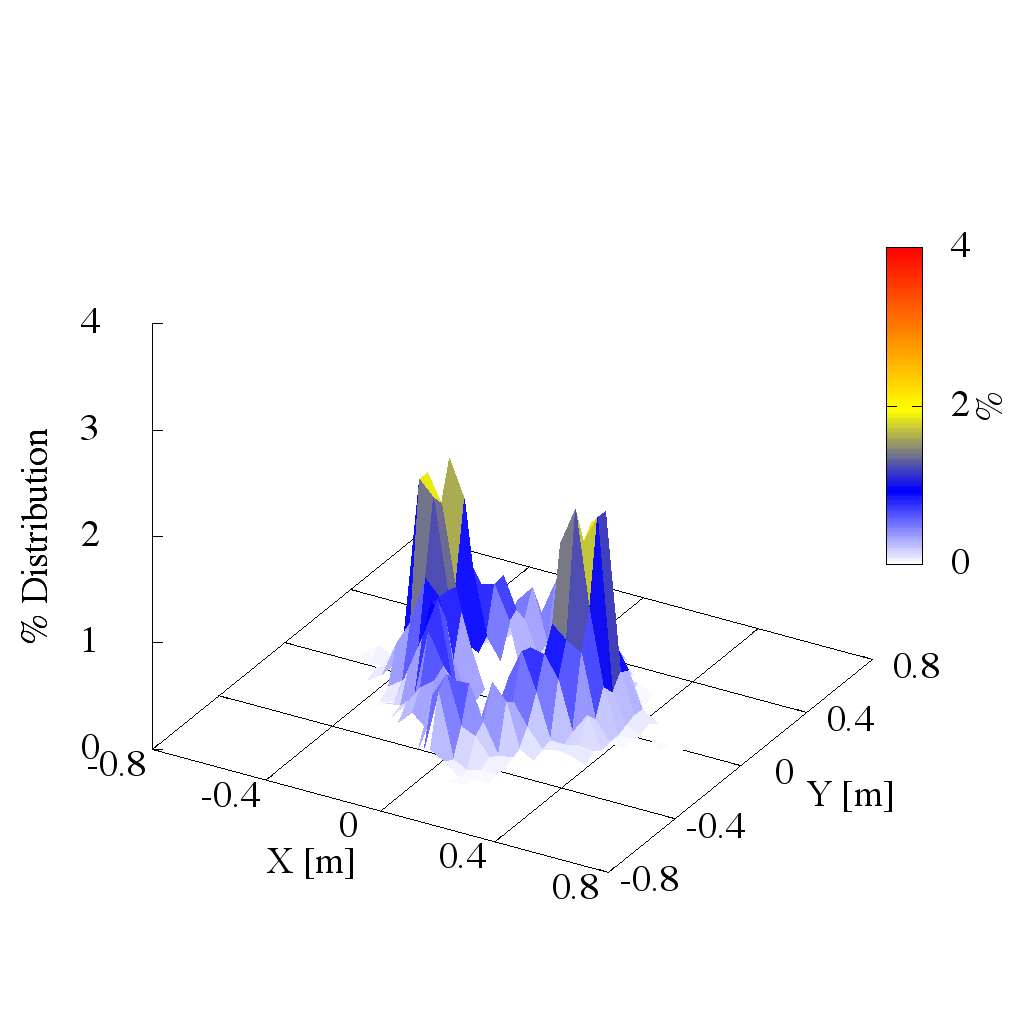}}
\subfigure[ Procedure No. 4 - Flow Ratio = .5]{\includegraphics[width=.43\textwidth]{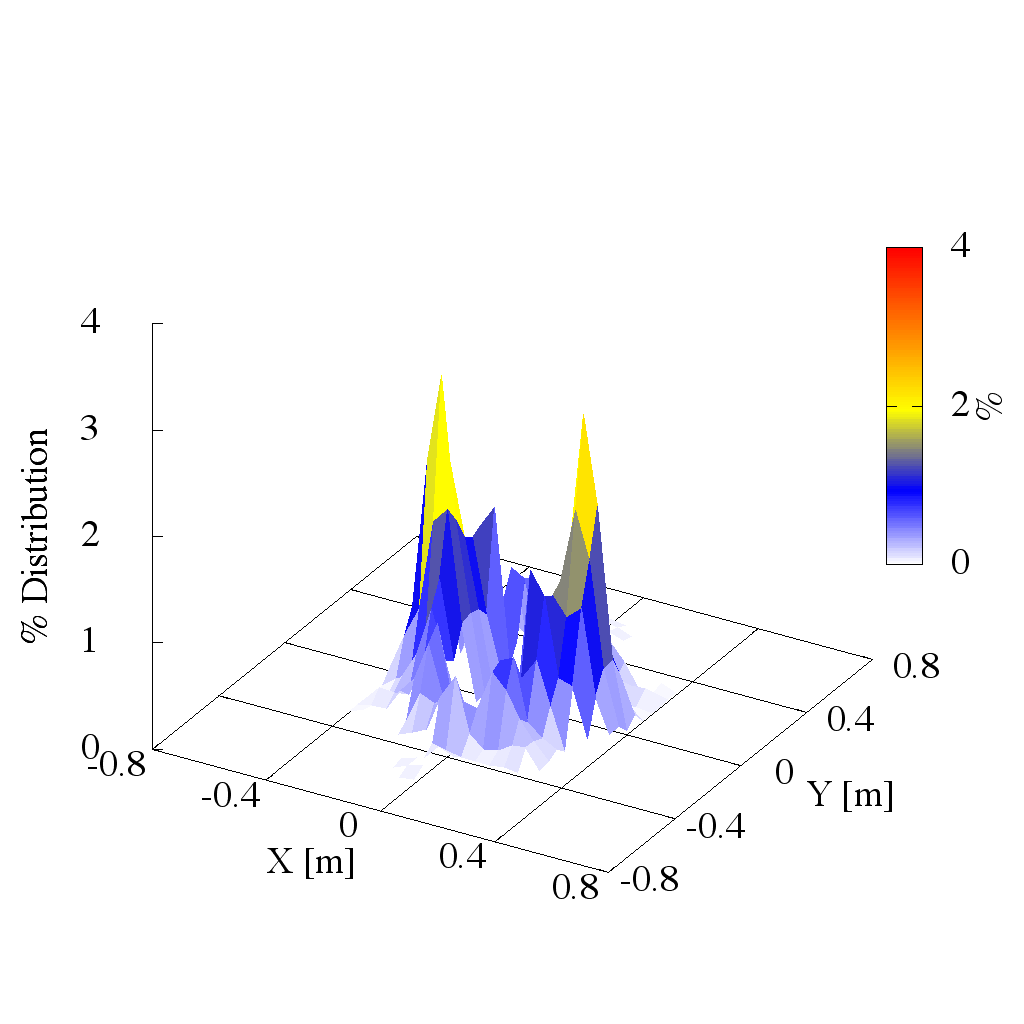}}
\caption{The percentage distribution of the relative positions of dyad members with respect to the centroid of the group. The distribution has been calculated with cells of .05 m side.}
\label{fig:pr}
\end{center}
\end{figure}

The distance of dyad members on the X-axis (.43 m $\pm$ .17 SD) and the Y-axis (.26 m $\pm$ .23 SD) were measured as the gap between the relative positions of members and the geometrical centre of the group (\emph{centroid}) (see Tab. \ref{tab:speed}, Fig. \ref{fig:comparison} and Fig. \ref{fig:pr}). A series of one factor analysis of variance (one-way ANOVA) showed a significant impact of the flow ratio on the X-distance of dyads [F(3,188) = 9.531, p $<$ .000]. A non significant impact of the flow ratio on the Y-distance was found [F(3,188) = 1.197, p value = .312]. A linear regression analysis showed a significant impact of the X-distance on the speed of dyads [F(1,190) = 32.180, p $<$ .000, with an R-square of .145; speed of dyads =  1.017 + .381 * X-distance]. A non significant impact was found for the Y-distance on speed [F(1,190) = 4.344, p value = .038, with an R-square of 0.022; speed of dyads =  1.165 - .150 * Y-distance]. 

\section{Conclusions}
The results of the experiment showed that Hp1 is partially confirmed: counter flow negatively impacted the speed of pedestrians. Considering bidirectional flow, the difference in the speeds of participants among procedure No. 3 and No. 4 was not significant. This result highlights that unbalanced flow ratio of .333 shows the same properties of balanced flow ratio, although separate studies came to partially different conclusions~\cite{PhysRevE.94.032304}.

Results about the impact of grouping showed that Hp2 is partially confirmed: in absence of counter flow the speeds of singles and dyads are not significantly different. In fact, in unidirectional flow dyad members walked freely with a line abreast pattern and an average Euclidean distance of .64 m $\pm$ .18 SD. In case of bidirectional flow, groups walked in average the 4\% slower than singles. The presence of a counter flow generally made dyad members to walk closer with an average Euclidean distance: .56 m $\pm$ .12 SD. Comparing to unidirectional flow, they had to modulate their relative positions from line abreast to diagonal and/or river-like patterns due to local situations of high density (see Fig. \ref{fig:pr}). Thus, bidirectional flow generated turbulences in the walking behaviour of dyads in terms of dynamic regulation of interpersonal distances and spatial pattern (proxemics), with detriment of speed.

This is confirmed by the analysis of trajectories among procedures (see Fig. \ref{fig:tr}): unidirectional procedure No. 1 is characterised by stable and  straight trajectories among all participants. Counter flow procedures describe instead more turbulent trajectories; however the level of density in the measurement area was sufficiently low to allow the formation of stable lanes. In conclusion, the results of the presented experiment could be of notable interest for those involved in the validation of pedestrian simulations, providing statistic models about speeds and spatial arrangement of dyads.  

\subsubsection*{Acknowledgement} 
The experiment was performed within the authorisation of The University of Tokyo. 

\bibliographystyle{IEEEtran}
\bibliography{template}

\end{document}